\documentclass[twocolumn,english]{revtex4-2}
\usepackage[T1]{fontenc}
\usepackage[latin9]{inputenc}
\usepackage{color}
\usepackage{amsmath}
\usepackage{graphicx}
\usepackage{esint}

\makeatletter
\usepackage{xcolor}
\usepackage{babel}
\usepackage{physics}

\makeatother

\usepackage{babel}
\begin{document}
\title{Neutral delay differential equation Kerr cavity model}
\author{Andrei G. Vladimirov and Daria A. Dolinina}
\affiliation{Weierstrass Institute, Mohrenstr. 39, 10117 Berlin, Germany}
\begin{abstract}
A neutral delay differential equation (NDDE) model of a Kerr cavity
with external coherent injection is developed that can be considered
as a generalization of the Ikeda map with second and higher order
dispersion being taken into account. It is shown that this model has
solutions in the form of dissipative solitons both in the limit, where
the model can be reduced to the Lugiato-Lefever equation (LLE), and
beyond this limit, where the soliton is eventually destroyed by the
Cherenkov radiation. Unlike the standard LLE the NDDE model is able
to describe the overlap of multiple resonances associated with different
cavity modes. 
\end{abstract}
\maketitle

\section{Introduction}

Over the last two decades, optical frequency combs have found applications
in different fields of science and industry including spectroscopy,
optical ranging, metrology, searching for exoplanets, microwave photonics
and optical communications \cite{cundiff2003colloquium,schroder2019laser,picque2019frequency,suh2019searching,trocha2018ultrafast,xue2016microwave}.
Standard methods of the frequency comb generation are based on the
use of mode-locked lasers \cite{jones2000carrier} and optical microcavities
subject to an external coherent injection \cite{del2007optical}.
One of the most commonly used methods to model Kerr optical cavities
is based on the application of the paradigmatic Lugiato-Lefever equation
(LLE) \cite{lugiato1987spatial}, which is known to exhibit S-shaped
branches of continuous wave (CW) solutions as well as dissipative
solitons preserving their shape in the course of propagation along
the cavity axis and sitting on a constant intensity background. The
latter solutions correspond to the \foreignlanguage{american}{so-called}
temporal cavity solitons (TCSs), which were observed experimentally
in optical microresonator frequency comb generators \cite{Herr14},
optical fiber cavities \cite{Leo10}, and mode-locked lasers \cite{grelu2012dissipative}.
The LLE based on the mean field approximation is, however, not free
from certain shortcomings. In particular, it describes bistable behavior
and TCS formation only in the vicinity of a single cavity resonance
\cite{Chembo2013,Herr14}. To overcome this shortcoming the modeling
approaches based on the so-called generalized Ikeda map \cite{mc1983solitary,haelterman1992dissipative,coen1997modulational,coen2013modeling,hansson2015frequency}
and a generalized LLE model with localized injection and losses \cite{kartashov2017multistability}
were developed to describe the appearance of the overlap of multiple
nonlinear resonances, multistability of CW solutions, \textcolor{black}{and
supersolitons}\textcolor{red}{{} }\textcolor{black}{\cite{hansson2015frequency}}.
Here we propose an alternative approach to model nonlinear dynamics
of an injected Kerr cavity based on second order neutral type delay
differential equation (NDDE), which also can be considered as a generalization
of the Ikeda map \cite{ikeda1979multiple}. Several DDE models were
already used previously to study the influence of chromatic dispersion
on the dynamics of laser systems \cite{heuck2010theory,pimenovprl,schelte2019third,schelte2020dispersive,pimenov20,pimenov2022temporal,seidel2022conservative}.
We show that in a certain limit the NDDE model can be reduced to the
LLE. We perform linear stability analysis of the NDDE model in the
practically important large delay limit and present numerical evidence
of the existence of stationary and oscillating dissipative optical
solitons in it. We show that beyond the LLE limit the TCSs of the
NDDE model are strongly affected by the Cherenkov radiation induced
by the high order dispersion. The modeling approach proposed here
can be adopted to study the dynamics of solid state and fiber lasers,
where the chromatic dispersion of the intracavity media plays an important
role in the mechanism of the short pulse generation as well as to
investigate the effect of second and higher order dispersion on the
characteristics of mode-locking regimes in semiconductor lasers. Furthermore,
the NDDE model might be useful to model the coupled-cavity systems
such as microcavity optical frequency comb generator pumped by a semiconductor
mode-locked light source, which is already successfully modeled by
DDE models \cite{VT05,vladimirov,VT04,viktorov2006model}.

\section{Model equation}

A neutral delay differential equation (NDDE) model of a ring Kerr
cavity subject to a coherent optical injection reads: 
\begin{gather}
\left(A+a\partial_{t}A+\frac{a^{2}-ib}{2}\partial_{tt}A\right)e^{-i\alpha\left|A\right|^{2}/2-i\theta/2}=\nonumber \\
\sqrt{\kappa}\left(A_{\tau}-a\partial_{t}A_{\tau}+\frac{a^{2}+ib}{2}\partial_{tt}A_{\tau}\right)e^{i\alpha\left|A_{\tau}\right|^{2}/2+i\theta/2}+\eta,\label{eq:Model}
\end{gather}
where $t$ is the time, $A(t)$ is the normalized electric field envelope,
$A_{\tau}=A\left(t-\tau\right)$ is the retarded field amplitude with
the delay time $\tau$ equal to the linear round-trip time of the
cavity. $\alpha$ is the Kerr coefficient, $\kappa$ is the linear
intensity attenuation factor per cavity round-trip, $\eta$ is the
normalized injection rate, and $\theta$ describes the detuning between
the injection frequency and the frequency of a cavity mode. The coefficients
$a>0$ and $b$ are responsible for the intracavity dispersion. Note
that, as it will be shown below, in the LLE limit only the coefficient
$b$ contributes to the second order dispersion, while the parameter
$a$ describes the group delay due to the first order dispersion.
In the absence of injection and losses, $\eta=0$ and $\kappa=1$,
Eq. (\ref{eq:Model}) is similar to the conservative version of the
Gires--Tournois interferometer model introduced in \cite{seidel2022conservative}.
Similarly to this model Eq. (\ref{eq:Model}) with $\eta=0$ and $\kappa=1$
is symmetric under the transformation $t\to-t$ and $A\to A^{*}$
combined with the time shift and, hence, it is reversible in the non-dissipative
limit. Note, however, that unlike the first order NDDE Gires--Tournois
interferometer model studied in \cite{seidel2022conservative} the
second order derivative terms responsible for the second order dispersion
in the LLE limit are present in Eq. (\ref{eq:Model}), see also Ref.
\cite{pimenov2022temporal}. This is an important difference between
Eq. (\ref{eq:Model}) and the model of Ref. \cite{seidel2022conservative}.
Note that for $a=b=0$ Eq. (\ref{eq:Model}) becomes similar to the
well known Ikeda map \cite{ikeda1979multiple}. The derivation of
Eq. (\ref{eq:Model}) is given in the next section together with the
derivation of three other versions of the NDDE model including the
mean field one. An advantage of the model equation (\ref{eq:Model})
is that in the non-dissipative limit, $\eta=0$ and $\kappa=1$, similarly
to the first order NDDE discussed in \cite{seidel2022conservative}
it admits a relatively simple conserved quantity $\partial_{t}W\left(t\right)=0$
with 
\begin{align*}
W\left(t\right) & =a^{3}\left|\partial_{t}A\right|^{2}+ib\left(A\partial_{t}A^{*}-A^{*}\partial_{t}A\right)+2a\left|A\right|^{2}\\
+ & \intop_{t}^{t+\tau}\left|A\left(x\right)+a\partial_{x}A\left(x\right)+\frac{a^{2}-ib}{2}\partial_{xx}A\left(x\right)\right|^{2}dx.
\end{align*}
Therefore this study is mainly focused on the analysis of Eq. (\ref{eq:Model}).

\section{Model derivation}

\begin{figure}
\includegraphics[scale=0.3]{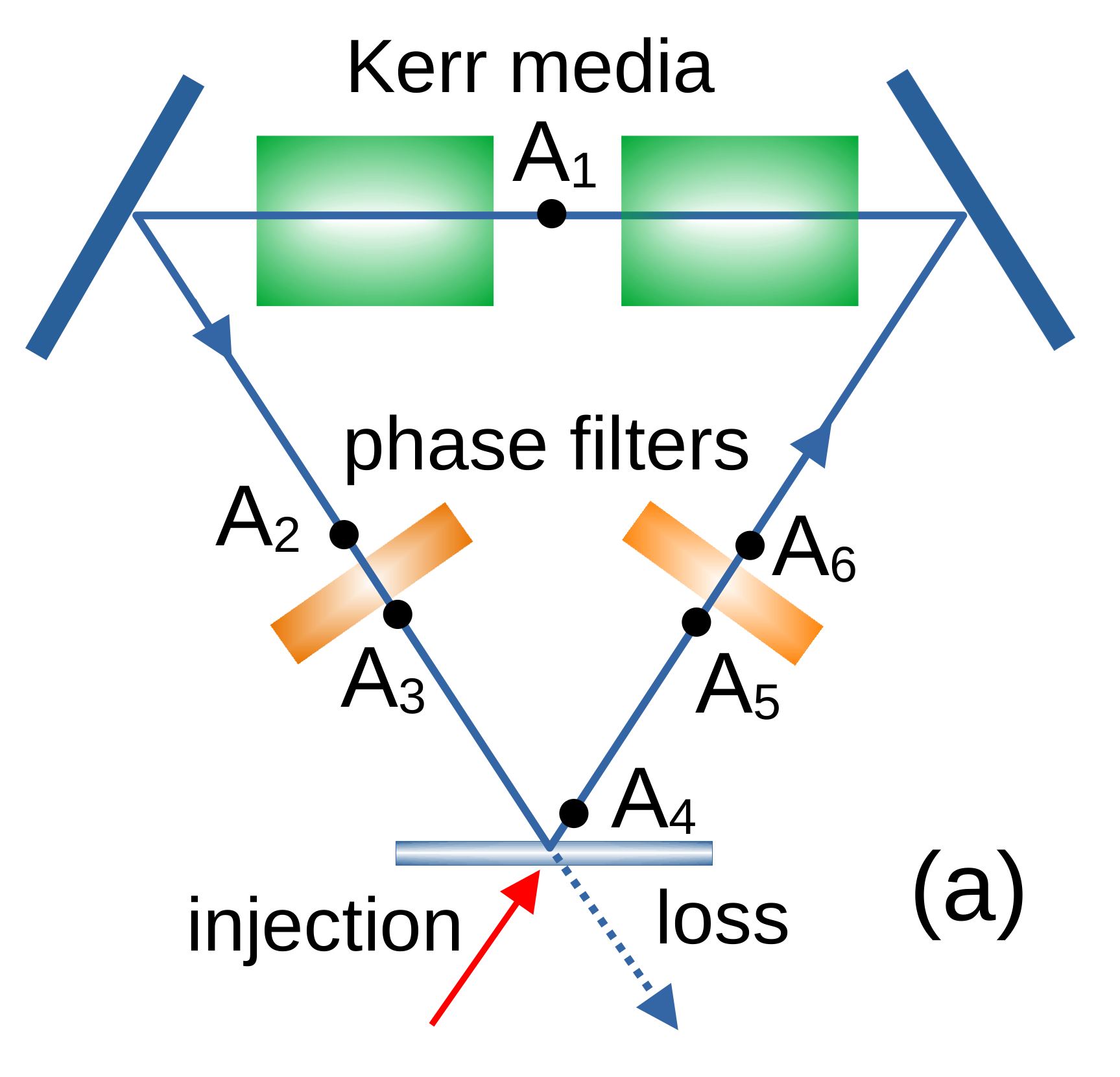}\includegraphics[scale=0.3]{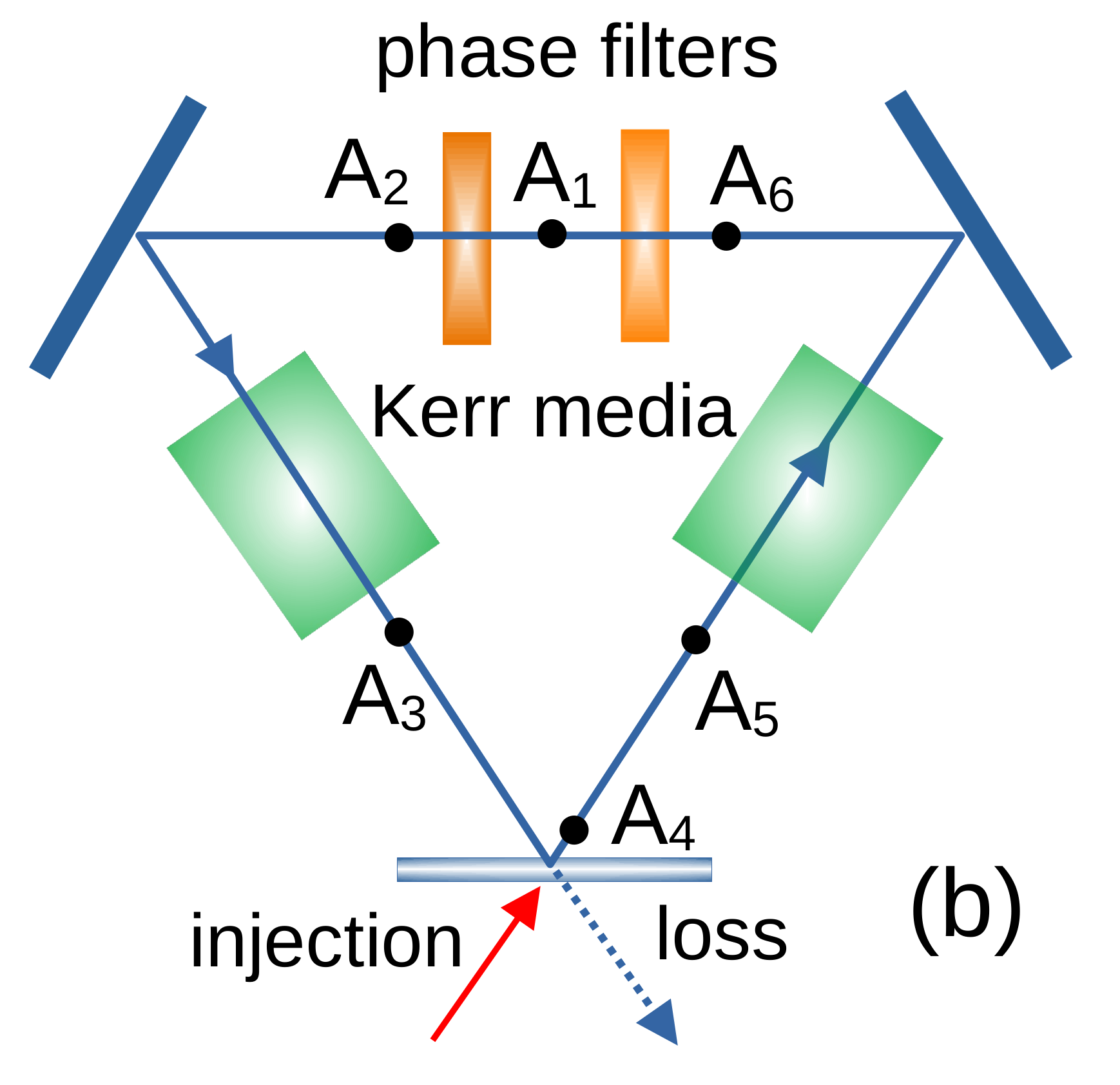}

\caption{Ring cavity with two Kerr media (green) and a pair of identical thin
dispersive elements (orange).\label{fig:Kerr-cavity-with}}
\end{figure}

To derive the NDDE model let us consider the schematic representation
of a ring Kerr cavity with a pair of thin dispersive elements and
two identical Kerr media shown in Fig. \ref{fig:Kerr-cavity-with}.
The field envelope on the output of the left Kerr medium is given
by $A_{2}\left(t+\tau_{1}\right)=A_{1}\left(t\right)e^{i\alpha\left|A_{1}\left(t\right)\right|^{2}/2+i\phi_{1}}$,
where $\alpha$ is the Kerr coefficient, $\phi_{1}$ and $\tau_{1}$
are the phase shift and the delay time due to the propagation in the
cavity, see Fig. \ref{fig:Kerr-cavity-with}(a).

The Fourier transform of the field envelope at the output from a thin
dispersive element (phase filter) is given by $\hat{A}_{3}\left(\omega\right)=\hat{f}\left(\omega\right)\hat{A}_{2}\left(\omega\right)$,
where $\hat{A}_{2}\left(\omega\right)$ is the Fourier transform of
the input field $A_{2}\left(t\right)$ and $\hat{f}\left(\omega\right)=e^{-i\Phi\left(\omega\right)}$
with real $\Phi\left(\omega\right)$, see e. g. Ref. \cite{heuck2010theory}.
Close to $\omega=0$ we can use the expansion 
\begin{gather}
\hat{f}\left(\omega\right)=e^{-i\Phi\left(0\right)}\nonumber \\
\times\left\{ 1+i\omega\Phi'\left(0\right)-\left[\Phi'\left(0\right)^{2}-i\Phi''\left(0\right)\right]\omega^{2}/2+{\cal O}\left(\omega^{3}\right)\right\} \label{eq:expansion}
\end{gather}
. Taking this relation into account, the electric field envelope $A_{3}$
can be expressed as 
\begin{gather}
A_{3}\left(t\right)={\cal F}^{-1}\left[\hat{f}\left(\omega\right)\hat{A}_{2}\left(\omega\right)\right]\nonumber \\
={\cal F}^{-1}\left\{ e^{-i\phi/2}\left[1+i\omega a-\frac{a^{2}+ib}{2}\omega^{2}+{\cal O}\left(\omega^{3}\right)\right]\hat{A}_{2}\left(\omega\right)\right\} \nonumber \\
=\left[1-a\partial_{t}+\frac{a^{2}+ib}{2}\partial_{tt}+{\cal O}\left(\partial_{ttt}\right)\right]A_{2}\left(t\right)e^{-i\phi/2},\label{eq:4}
\end{gather}
where ${\cal F}^{-1}$ is the inverse Fourier transform and the notations
$\Phi\left(0\right)\equiv\phi$, $\Phi'\left(0\right)\equiv a$, and
$\Phi''\left(0\right)\equiv-b$ are used. The field amplitude $A_{4}$
is obtained from $A_{3}$ by taking into account time delay $\tau_{2}$
, phase shift $\phi_{2}$, and introducing the injection rate $\eta$,
and the intensity attenuation factor $\kappa$ due to the cavity losses.
Thus we get 
\begin{gather}
A_{4}\left(t+\frac{\tau}{2}\right)=\sqrt{\kappa}\left[1-a\partial_{t}\right.\nonumber \\
\left.+\frac{a^{2}+ib}{2}\partial_{tt}+{\cal O}\left(\partial_{ttt}\right)\right]\left(A_{1}e^{i\alpha\left|A_{1}\right|^{2}/2+i\theta/2}\right)+\eta,\label{eq:A3PT}
\end{gather}
with the one half of the cavity round trip time $\tau/2=\tau_{1}+\tau_{2}$
and the phase shift $\theta/2=-\left(\phi_{1}+\phi+\phi_{2}\right)$.

The field envelope on the input of the left Kerr medium is given by
$A_{6}\left(t-\tau_{1}\right)=A_{1}\left(t\right)e^{-i\alpha\left|A_{1}\left(t\right)\right|^{2}/2-i\phi_{1}}$
with $\left|A_{1}\left(t\right)\right|^{2}=\left|A_{6}\left(t-\tau_{1}\right)\right|^{2}$.
The Fourier transform of the field envelope $A_{6}$ is given by $\hat{A}_{6}\left(t\right)=\hat{f}\left(\omega\right)\hat{A}_{5}\left(\omega\right)$,
which can be rewritten as $\hat{A}_{5}\left(t\right)=\hat{f}^{-1}\left(\omega\right)\hat{A}_{6}\left(\omega\right)=\hat{f}^{*}\left(\omega\right)\hat{A}_{6}\left(\omega\right)$.
Hence we get 
\begin{eqnarray*}
A_{5}\left(t\right) & = & {\cal F}^{-1}\left[\hat{f}^{*}\left(\omega\right)\hat{A}_{6}\left(\omega\right)\right]=
\end{eqnarray*}
\[
=\left[1+a\partial_{t}+\frac{a^{2}-ib}{2}\partial_{tt}+{\cal O}\left(\partial_{ttt}\right)\right]A_{6}\left(t\right)e^{-i\phi/2}.
\]
Therefore, similarly to (\ref{eq:A3PT}) we obtain 
\begin{gather}
A_{4}\left(t-\frac{\tau}{2}\right)=\left[1+a\partial_{t}\right.\nonumber \\
\left.+\frac{a^{2}-ib}{2}\partial_{tt}+{\cal O}\left(\partial_{ttt}\right)\right]\left(A_{1}e^{-i\alpha\left|A_{1}\right|^{2}/2-i\theta/2}\right).\label{eq:A3}
\end{gather}

Finally, shifting the time in Eq. (\ref{eq:A3PT}) by $-\tau$, equating
the resulting equation to Eq. (\ref{eq:A3}), and neglecting high
order dispersion terms ${\cal O}\left(\partial_{ttt}\right)$ we get
\[
\left(1+a\partial_{t}+\frac{a^{2}-ib}{2}\partial_{tt}\right)\left(Ae^{-i\alpha\left|A\right|^{2}/2-i\theta/2}\right)=
\]
\begin{equation}
\sqrt{\kappa}\left(1-a\partial_{t}+\frac{a^{2}+ib}{2}\partial_{tt}\right)\left(A_{\tau}e^{i\alpha\left|A_{\tau}\right|^{2}/2+i\theta/2}\right)+\eta,\label{eq:Model-2}
\end{equation}
where $A\equiv A_{1}$ and $A_{\tau}=A\left(t-\tau\right)$.

The model (\ref{eq:Model}) can be derived in a similar way to Eq.
(\ref{eq:Model-2}), see Fig. \ref{fig:Kerr-cavity-with}(b). Here
neglecting the high order dispersions we get:

\begin{equation}
A_{2}\approx A_{1}-a\partial_{t}A_{1}+\frac{a^{2}+ib}{2}\partial_{tt}A_{1},\label{eq:A2}
\end{equation}
and 
\begin{equation}
A_{4}\left(t+\frac{\tau}{2}\right)\approx\sqrt{\kappa}A_{2}e^{i\alpha\left|A_{2}\right|^{2}/2+i\theta/2}+\eta.\label{eq:A4}
\end{equation}
Substituting Eq. (\ref{eq:A2}) into Eq. (\ref{eq:A4}), assuming
that $a$, $b$, and $\alpha$ are sufficiently small, and neglecting
the ${\cal O}\left(\alpha a\right)$ and ${\cal O}\left(\alpha b\right)$
terms in the exponential we obtain 
\begin{gather}
A_{4}\left(t+\frac{\tau}{2}\right)\approx\nonumber \\
\sqrt{\kappa}\left(A_{1}-a\partial_{t}A_{1}+\frac{a^{2}+ib}{2}\partial_{tt}A_{1}\right)e^{i\alpha\left|A_{1}\right|^{2}/2+i\theta/2}+\eta.\label{eq:A41}
\end{gather}
Similarly, for another half of the cavity, we get 
\begin{gather}
A_{4}\left(t-\frac{\tau}{2}\right)\approx\nonumber \\
\left(A_{1}+a\partial_{t}A_{1}+\frac{a^{2}-ib}{2}\partial_{tt}A_{1}\right)e^{-i\alpha\left|A_{1}\right|^{2}/2-i\theta/2}.\label{eq:A42}
\end{gather}
Finally, combining Eqs. (\ref{eq:A41}) and (\ref{eq:A42}) we arrive
at the NDDE model (\ref{eq:Model}). The latter equation can be further
simplified under the approximation $\alpha a,\alpha b\ll1$, which
was already used above:

\begin{gather}
\begin{gathered}\left(Ae^{-i\alpha\left|A\right|^{2}/2}+a\partial_{t}A+\frac{a^{2}-ib}{2}\partial_{tt}A\right)e^{-i\theta/2}\\
=\sqrt{\kappa}\left(A_{\tau}e^{i\alpha\left|A\right|^{2}/2}-a\partial_{t}A_{\tau}+\frac{a^{2}+ib}{2}\partial_{tt}A_{\tau}\right)e^{i\theta/2}+\eta,
\end{gathered}
\label{eq:Model-3}
\end{gather}
In the mean field approximation, where the field amplitude is small,
expanding the exponentials in Eq. (\ref{eq:Model-3}) we get the following
equation: 
\begin{gather}
\left(A+a\partial_{t}A+\frac{a^{2}-ib}{2}\partial_{tt}A-i\frac{\alpha}{2}A\left|A\right|^{2}\right)e^{-i\theta/2}\nonumber \\
=\sqrt{\kappa}\left(A_{\tau}-a\partial_{t}A_{\tau}+\frac{a^{2}+ib}{2}\partial_{tt}A_{\tau}+i\frac{\alpha}{2}A_{\tau}\left|A_{\tau}\right|^{2}\right)e^{i\theta/2}+\eta.\label{eq:Model-mean_field}
\end{gather}

Neutral DDEs (\ref{eq:Model}), (\ref{eq:Model-2}), (\ref{eq:Model-3}),
and (\ref{eq:Model-mean_field}) are reversible in the non-dissipative
limit $\eta=0$ and $\kappa=1$. This property is similar to that
of the LLE, which will be derived from these equations in the next
section. As it will be shown in the next section, the parameter $b$
corresponds to the second order dispersion coefficient in the LLE
limit. In the absence of dispersion according to Eq. (\ref{eq:A2})
the field amplitude on the output of the dispersive element is defined
by 
\[
A_{2}\approx A_{1}-a\partial_{t}A_{1}+\frac{a^{2}}{2}\partial_{tt}A_{1}=A_{1}\left(t-a\right)+{\cal O}\left(a^{3}\right).
\]
Therefore, the parameter $a$ has the meaning of the group delay introduced
by the dispersive element. The inequality $a>0$ follows from the
causality principle. As it will be shown below this inequality is
the necessary but not sufficient condition of the absence of a spurious
instability in the NDDE models.

Note, that a straightforward derivation of the model equation without
splitting Kerr and dispersive media into two symmetric parts would
result in a ``regular'' DDE model instead of NDDE: 
\begin{gather}
A+2a\partial_{t}A+\left(2a^{2}-ib\right)\partial_{tt}A\nonumber \\
=\sqrt{\kappa}A\left(t-\tau\right)e^{i\alpha\left|A\left(t-\tau\right)\right|^{2}+i\theta}+\eta,\label{eq:Pim}
\end{gather}
which is similar to the generalization of the Ikeda map derived in
\cite{pimenov2022temporal} to describe a Kerr cavity with two spectral
filters. After rescaling the time variable $t\to\sqrt{2}at$ the DDE
model (\ref{eq:Pim}) is transformed into Eq. (12) of Ref. \cite{pimenov2022temporal}
with the constraint $\sigma=\sqrt{2}$. This constraint, however,
contradicts to the applicability condition of the DDE model of Ref.
\cite{pimenov2022temporal}. Below, in Sec. \ref{sec:Linear-stability-analysis}
we show that Eq. (\ref{eq:Pim}) with $b\neq0$ demonstrates a spurious
instability in the zero loss and injection limit, $\kappa=1$ and
$\eta=0$. This instability is related to a dissipation introduced
by the truncation of the series expansion in Eq. (\ref{eq:expansion}).
On the other hand, in the derivation of the NDDE model (\ref{eq:Model})
we have used two truncations of the two transfer functions and in
this case the corresponding dissipative contributions cancel each
other. 

To conclude this section, we note that the derivation presented here
can be trivially generalized by including higher order dispersion
terms into the model equation.

\section{Reduction to the LLE}

Using the multiscale method described in \cite{kolokolnikov2006q}
we can derive the same LLE from any one of the four model equations
(\ref{eq:Model}), (\ref{eq:Model-2}), (\ref{eq:Model-3}), and (\ref{eq:Model-mean_field}).
To be more specific, we choose Eq. (\ref{eq:Model}) and in the large
delay limit $\tau=\epsilon^{-1}\gg1$ rescale the time as $x=\epsilon t$
to get 
\[
\left(A+\epsilon a\partial_{x}A+\epsilon^{2}\frac{a^{2}-ib}{2}\partial_{xx}A\right)e^{i\alpha\left|A\right|^{2}/2+i\theta/2}=
\]
\begin{equation}
\sqrt{\kappa}\left(A_{\tau_{1}}-\epsilon a\partial_{x}A_{\tau_{1}}+\epsilon^{2}\frac{a^{2}+ib}{2}\partial_{xx}A_{\tau_{1}}\right)e^{i\alpha\left|A_{\tau_{1}}\right|^{2}/2+i\theta/2}+\eta,\label{eq:Model-rescaled}
\end{equation}
where $A_{\tau_{1}}=A\left(t-1\right)$. Next, we introduce two time
scales $A\left(t\right)=u\left(t_{0},t_{2}\right)$, where $t_{0}=\left(1-\epsilon c_{1}+\epsilon^{2}c_{2}+\dots\right)x$
, $t_{2}=\epsilon^{2}x$, and rescale the parameters as 
\begin{equation}
\alpha=\epsilon^{2}\chi,\quad\kappa=e^{-2\epsilon^{2}k},\quad\eta=\epsilon^{2}r.\label{eq:LLE_limit1}
\end{equation}
Substituting these expressions into Eq. (\ref{eq:Model-rescaled}),
collecting the first order terms in $\epsilon$, and using the Fredholm
alternative we obtain the following boundary condition 
\begin{equation}
u\left(t_{0},t_{2}\right)=u\left(t_{0}-1,t_{2}\right)e^{i\theta},\label{eq:bc_lle}
\end{equation}
together with the relation $c_{1}=2a$ for the group delay parameter.
Finally, collecting the second order terms in $\epsilon$ we get $c_{2}=4a^{2}$
and the LLE model: 
\begin{equation}
\partial_{t_{2}}u=-ku+i\chi u\left|u\right|^{2}+ib\partial_{t_{0}t_{0}}u+r.\label{eq:LLE-generalized}
\end{equation}
Note that Eq. (\ref{eq:LLE-generalized}) with the boundary condition
(\ref{eq:bc_lle}), where the parameter $\theta$ plays the role of
the detuning, is able to describe multiple resonances corresponding
to different cavity modes. However, it cannot describe the overlap
of these resonances. In order to get overlapping resonances and coexisting
different solitons one needs to replace distributed injection and
losses with the localized ones, as it was proposed in \cite{kartashov2017multistability}.
It is seen from Eq. (\ref{eq:LLE-generalized}) that the coefficient
$b$ in front of the second derivative terms is responsible for the
second order dispersion in the LLE limit, see also \cite{pimenov2022temporal}.
Therefore, these derivatives cannot be neglected in the model equation. 

Using the additional assumption that the detuning is small, 
\begin{equation}
\theta=-\epsilon^{2}\Theta,\label{eq:LLE_limit2}
\end{equation}
instead of (\ref{eq:LLE-generalized}) we get the standard LLE describing
a vicinity of a single cavity resonance 
\begin{equation}
\partial_{t_{2}}u=-ku-i\Theta u+i\chi u\left|u\right|^{2}+ib\partial_{t_{0}t_{0}}u+r\label{eq:LLE-standard}
\end{equation}
with the periodic boundary condition $u\left(t_{0},t_{2}\right)=u\left(t_{0}-1,t_{2}\right)$.

Note that, unlike Eqs. (\ref{eq:LLE-generalized}) and (\ref{eq:LLE-standard})
obtained by collecting $\epsilon^{2}$-terms, in Ref. \cite{seidel2022conservative},
where the nonlinear Schr{ö}dinger equation (NLSE) was derived from
a NDDE without second order derivatives and the external injection,
the second and third order dispersion terms appear simultaneously
in the oder $\epsilon^{3}$. The NLSE is then obtained by choosing
the detuning in such a way that the third order dispersion vanishes.
In the conservative limit the LLE (\ref{eq:LLE-standard}) also transforms
into the NLSE. However, the existence of stable soliton solutions
is hardly possible in the full NDDE model near this limit. Since unlike
the standard NLSE the NDDE contains the dispersion of all the orders,
the next order after the second order dispersion will be the third
order one. However, it is well known that the solitons of the NLSE
are destroyed by the Cherenkov radiation \cite{Karpman93,Akhmediev95}
in the presence of an arbitrary small third order dispersion term
\cite{Wai90,elgin1992soliton,gordon1992dispersive}.

\section{Linear stability analysis in the small amplitude conservative limit\label{sec:Linear-stability-analysis}}

\begin{figure}

\includegraphics[scale=0.5]{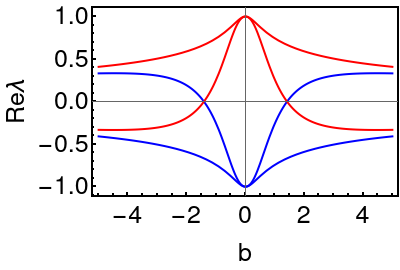}

\caption{Real parts of the eigenvalues $\lambda_{\pm}$ defined by (\ref{eq:Eig})
with $a=1$ (blue) and $a=-1$ (red).\label{fig:Real-parts}}
\end{figure}

Let us consider the linear equation 
\begin{gather}
A+a\partial_{t}A+\frac{a^{2}-ib}{2}\partial_{tt}A\nonumber \\
=\left(A_{\tau}-a\partial_{t}A_{\tau}+\frac{a^{2}+ib}{2}\partial_{tt}A_{\tau}\right)e^{i\theta},\label{eq:Model_lin}
\end{gather}
obtained by the linearization of the NDDE models with $\kappa=1$
and $\eta=0$ at the trivial solution $A=0.$ Substituting $A\left(t\right)=A_{0}e^{\lambda t}$
into (\ref{eq:Model_lin}) we get the characteristic equation 
\begin{equation}
1+a\lambda+\frac{a^{2}-ib}{2}\lambda^{2}=\left(1-a\lambda+\frac{a+ib}{2}\lambda^{2}\right)e^{-\lambda\tau}.\label{eq:Char}
\end{equation}
In the large delay limit using the approach of Ref. \cite{Yanchuk2010a}
we substitute $e^{-\lambda\tau}\equiv Y$ into Eq. (\ref{eq:Char}),
solve the resulting equation with respect to $Y$, and set $\lambda=i\nu$.
Thus, we get the following expression for the pseudo-continuous spectrum
in the limit $\tau\to\infty$ 
\begin{gather}
\lambda\tau=-\ln\left[\frac{1+ia\nu-\left(a^{2}-ib\right)\nu^{2}/2}{1-ia\nu-\left(a^{2}+ib\right)\nu^{2}/2}e^{-i\theta}\right]=\nonumber \\
-\ln e^{i\Psi\left(\nu\right)-i\theta}=-i\Psi\left(\nu\right)+i\theta\label{eq:lambda}
\end{gather}
with real $\Psi\left(\nu\right)$. The fact that the pseudo-continuous
spectrum of Eq. (\ref{eq:Model_lin}) is purely imaginary is obviously
the consequence of the reversibility of this equation, see also Ref.
\cite{seidel2022conservative}. However, whether this system exhibits
spurious instability depends on the discrete spectrum of the characteristic
equation (\ref{eq:Char}). In the large delay limit discrete eigenvalues
with positive real parts correspond to small $\left|Y\right|\ll1$
in the limit $\tau\to\infty$. Therefore, they can be obtained by
equating to zero the left hand side of Eq. (\ref{eq:Char}), $1+a\lambda+(a^{2}-ib)\lambda^{2}/2=0$,
which gives 
\begin{equation}
\lambda_{\pm}=\frac{-a\pm\sqrt{-a^{2}+2ib}}{a^{2}-ib}.\label{eq:Eig}
\end{equation}
The real parts of the eigenvalues $\lambda_{\pm}$ are shown in Fig.
\ref{fig:Real-parts} as functions of the ``second order dispersion
coefficient'' $b$. It is seen that for $a<0$ one of the two discrete
eigenvalues always has a positive real part. This means that Eq. (\ref{eq:Model_lin})
with $a<0$ should demonstrate a spurious instability. For $a>0$
both the eigenvalues $\lambda_{\pm}$ have negative real parts if
\begin{equation}
\left|b\right|<\sqrt{2}a^{2}.\label{eq:ineq}
\end{equation}
Hence, in order to avoid spurious instabilities below we choose $a>0$
in the NDDE models and assume that the inequality (\ref{eq:ineq})
is satisfied. Note that positive coefficient\textbf{ $a$} can be
rescaled to unity by rescaling the time variable in the model equations.

Substituting $\kappa=1$ and $\eta=0$ into Eq. (\ref{eq:Pim}) and
linearizing this equation at $A=0$, instead of (\ref{eq:lambda})
we get $\lambda\tau=-\ln\left[1+2ia\nu-\left(2a^{2}-ib\right)\nu^{2}\right]+i\theta$.
In particular, at $\nu=0$ one obtains 
\[
\Re\lambda=\Re\frac{d\lambda}{d\nu}=\Re\frac{d^{2}\lambda}{d\nu^{2}}=0,\quad\Re\frac{d^{3}\lambda}{d\nu^{3}}=-\frac{3b}{\tau}.
\]
It follows from these relations that for any nonzero $b$ and sufficiently
small $\nu$ the curve of the pseudo-continuous spectrum visits the
right halfplane of the complex plane. Therefore, in this case one
can expect an instability associated with the pseudo-continuous spectrum
in the large delay limit.

\section{CW solutions}

\begin{figure}[t]
\includegraphics[scale=0.4]{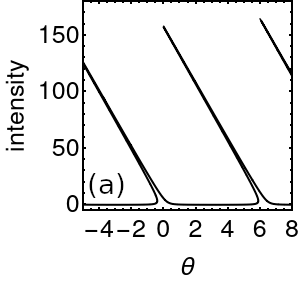}\includegraphics[scale=0.39]{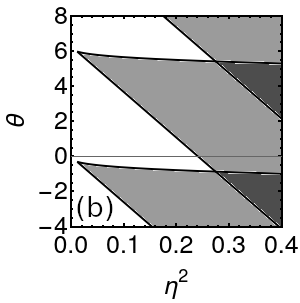}

\caption{Intensity of CW solitions of Eq. (\ref{eq:Model}) as a function of
the detuning parameter $\theta$ for $\eta=0.5$ (a). Saddle-node
bifurcations of CW solutions on ($\eta^{2},\text{\ensuremath{\theta}}$)-plane
(b). White, light gray, and dark gray areas limited by the saddle-node
bifurcations indicate the existence of one, three, and five CW solutions,
respectively. The parameter values $\kappa=0.923$ and $\alpha=0.04$
correspond to $\epsilon=0.2$ in Eq. (\ref{eq:LLE_limit1}). \label{fig:CW_intensity} }
\end{figure}

In this section we study the linear stability of the CW solutions
of Eq. (\ref{eq:Model}). The amplitude $A$ of the CW solutions satisfies
the equation 
\begin{gather}
Ae^{-i\alpha\left|A\right|^{2}/2-i\theta/2}=\sqrt{\kappa}Ae^{i\alpha\left|A\right|^{2}/2+i\theta/2}+\eta.\label{eq:cw}
\end{gather}
Substituting into Eq. (\ref{eq:cw}) $A=\sqrt{I}e^{i\varphi}$, where
$I$ and $\varphi$ are the intensity and the phase of the CW solution,
and separating real and imaginary parts of the resulting equation
we get the following equation for the CW intensity 
\begin{equation}
\left[1+\kappa-2\sqrt{\kappa}\cos\left(\theta+\alpha I\right)\right]I=\eta^{2},\label{eq:cwcond}
\end{equation}
which is transformed into the corresponding relation of the LLE in
the limit (\ref{eq:LLE_limit1}) and (\ref{eq:LLE_limit2}). The phase
of the CW solution is given by 
\[
\tan\varphi=\frac{\sin\left(\frac{\alpha I}{2}\right)+\sqrt{\kappa}\sin\left(\theta+\frac{\alpha I}{2}\right)}{\cos\left(\frac{\alpha I}{2}\right)-\sqrt{\kappa}\cos\left(\theta+\frac{\alpha I}{2}\right)}.
\]
Nonlinear resonances of CW solutions of Eq. (\ref{eq:Model}) near
cusp bifurcations corresponding to different cavity modes are shown
in Fig. \ref{fig:CW_intensity}(a).

The saddle-node bifurcations of the CW solutions are defined by the
conditions

\[
4\alpha^{2}\kappa I^{4}-\alpha^{2}I^{2}\left[\text{\ensuremath{\eta^{2}}}-\left(\kappa+1\right)I\right]^{2}=\text{\ensuremath{\eta}}^{4}
\]

\begin{gather*}
\theta_{\pm}=\arctan\left[\frac{-\left(1+\kappa\right)\alpha I\pm\sqrt{\kappa\left(4\alpha^{2}I^{2}+2-\kappa\right)-1}}{1+\kappa\pm\alpha I\sqrt{\kappa\left(4\alpha^{2}I^{2}+2-\kappa\right)-1}}\right]\\
-\alpha I+2\pi n,
\end{gather*}
where $\theta_{\pm}$ with $n=0,\pm1,\pm2\dots$ define pairs of saddle-node
bifurcation curves originating from the cusp bifurcation points corresponding
to different cavity resonances. Figure \ref{fig:CW_intensity}(b)
illustrates the coexistence of multiple CW solutions of Eq. (\ref{eq:Model}).
It is seen that at sufficiently high injection rates the CW resonances
start overlapping. Unlike the resonances shown in Fig. \ref{fig:CW_intensity}(b),
the resonances of Eq. (\ref{eq:Model-mean_field}) with cubic nonlinearity
are much more narrow and never overlap. This can be easily understood
by noticing that for cubic nonlinearity Eq. (\ref{eq:cwcond}) transforms
into a cubic equation in $I$ having not more than three solutions.
Hence, the mean field model (\ref{eq:Model-mean_field}) does not
describe well the dynamics of Eq. (\ref{eq:Model}) beyond the LLE
limit. Therefore, below we focus mainly on the model (\ref{eq:Model}),
which does not assume the mean field approximation.

To study the linear stability of the CW solution we linearize Eq.
(\ref{eq:Model}) around this solution, $A(t)=\left(\sqrt{I}+\delta Ae^{\lambda t}\right)e^{i\varphi}$.
Thus we get the following characteristic equation: 
\begin{equation}
c_{2}Y^{2}+c_{1}Y+c_{0}=0,\label{Y_eig}
\end{equation}
where $Y(\lambda)=\exp(-\lambda\tau)$, and the coefficients $c_{0,1,2}$
are polynomoals in $\lambda$: 
\[
c_{2}=\kappa+\frac{\kappa\lambda}{4}\left[\left(a^{4}+b^{2}\right)\lambda^{3}-4a^{3}\lambda^{2}+2\left(4a^{2}-b\alpha I\right)\lambda-8a\right],
\]
\begin{equation}
\begin{aligned}c_{1}= & \frac{\sqrt{\kappa}}{2}\left\{ \left[\left(b^{2}-a^{4}\right)\lambda^{4}-2b\alpha I\lambda^{2}-4\right]\cos\left(\alpha I+\theta\right)\right.\\
 & \left.+2(2+a^{2}\lambda^{2})(\alpha I-b\lambda^{2})\sin\left(\alpha I+\theta\right)\right\} 
\end{aligned}
\label{eq:c2-0}
\end{equation}
\[
c_{0}=\frac{1}{4}\left[\left(a^{4}+b^{2}\right)\lambda^{4}+4a^{3}\lambda^{3}+2\left(4a^{2}-b\alpha I\right)\lambda^{2}+8a\lambda+4\right].
\]
In the limit of large delay time $\tau\rightarrow\infty$ the eigenvalues
of the pseudo-continuous spectrum (PCS) can be represented as $\lambda\approx i\mu+\gamma/\tau$
with real $\mu$ and $\gamma$ \cite{Yanchuk2010a}. The pseudo-continuous
spectrum is given by the two solution branches of the quadratic equation
(\ref{Y_eig}): 
\begin{equation}
\gamma\left(\mu\right)=-\Re\left\{ \ln\left[Y\left(i\mu\right)\right]\right\} .
\end{equation}
Stable CW solutions are characterized by $\gamma\left(\mu\right)<0$
and, in particular, $\gamma\left(0\right)<0$. At the saddle-node
(flip) bifurcation point we have $\gamma\left(0\right)=0$ and $Y\left(0\right)=1$
($\gamma\left(0\right)=0$ and $Y\left(0\right)=-1$), while modulational
instability takes place when one or both the branches of pseudo-continuous
spectrum are tangent to the imaginary axis at $\mu=\pm\mu_{m}$ with
$\left|\mu_{m}\right|>0$.

\begin{figure}[t]
\includegraphics[width=0.5\linewidth]{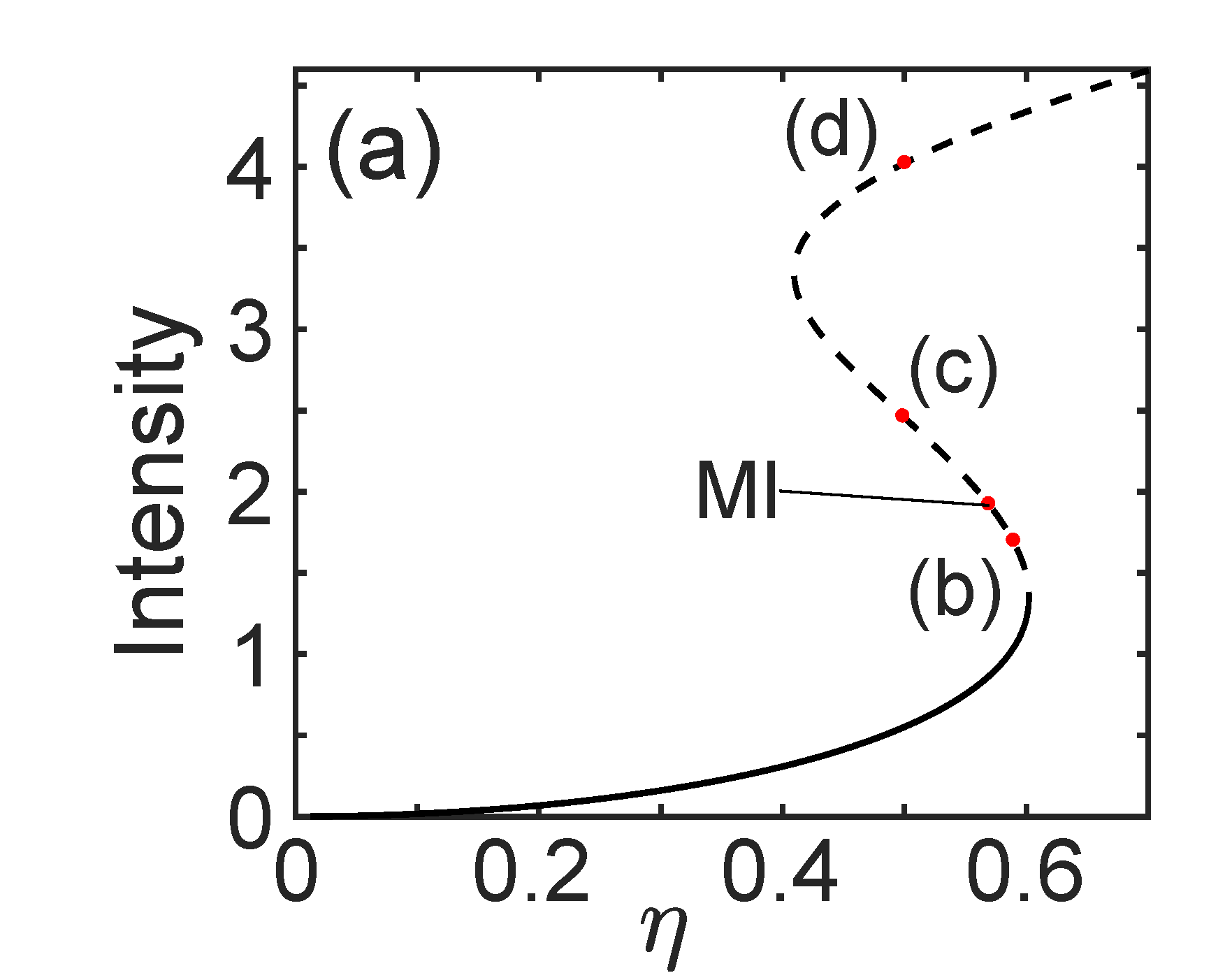}\includegraphics[width=0.5\linewidth]{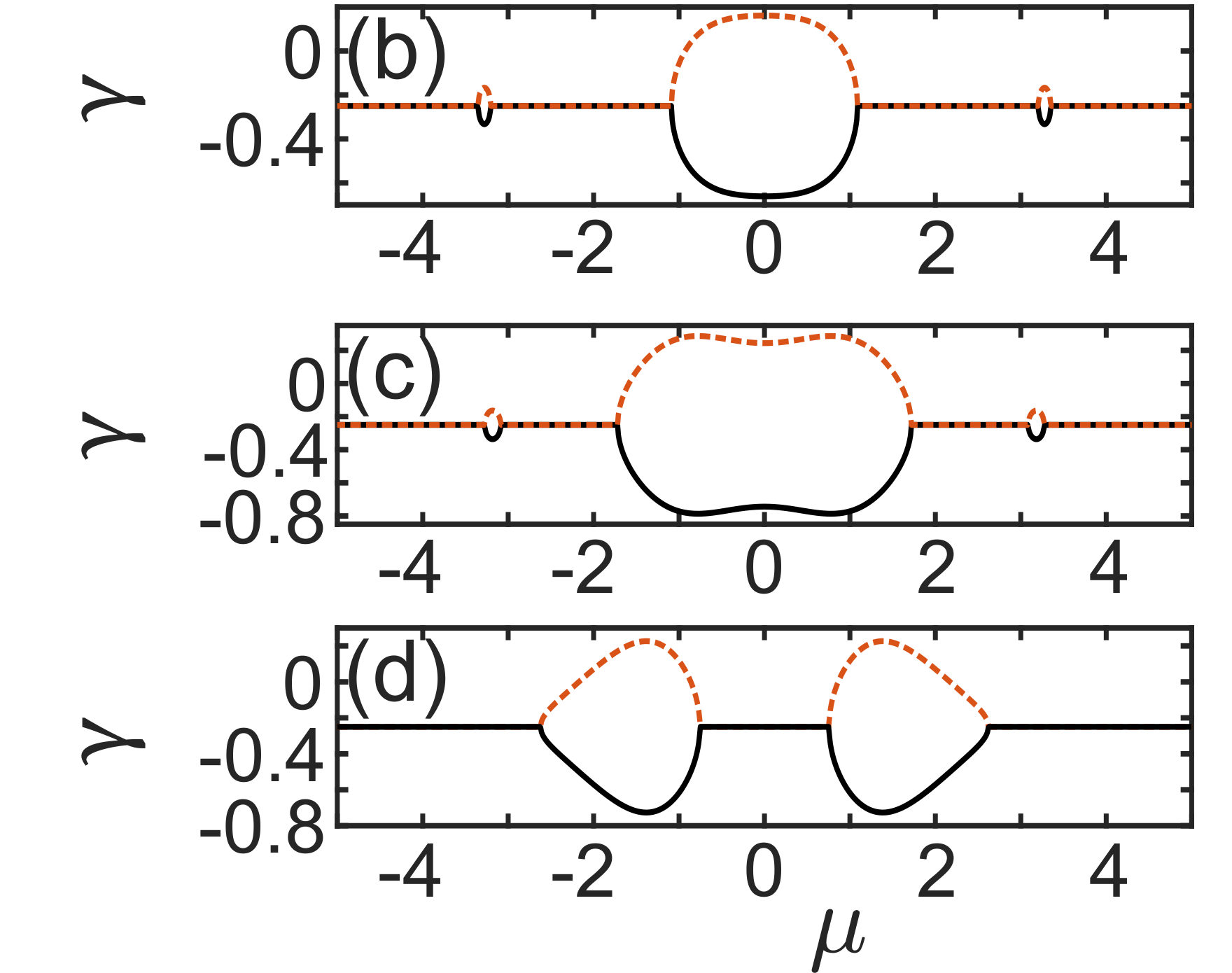}\\
 \includegraphics[scale=0.3]{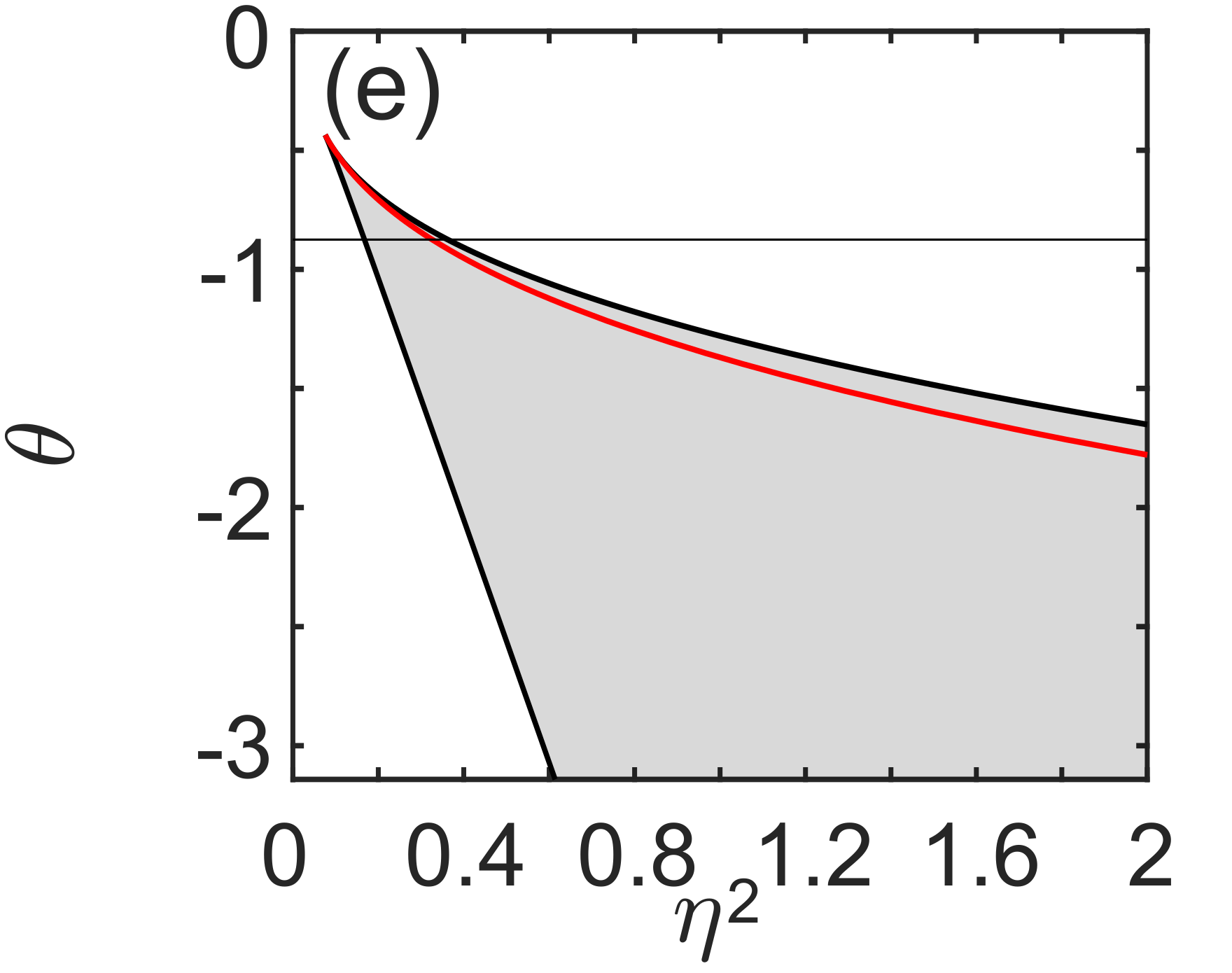}\includegraphics[scale=0.3]{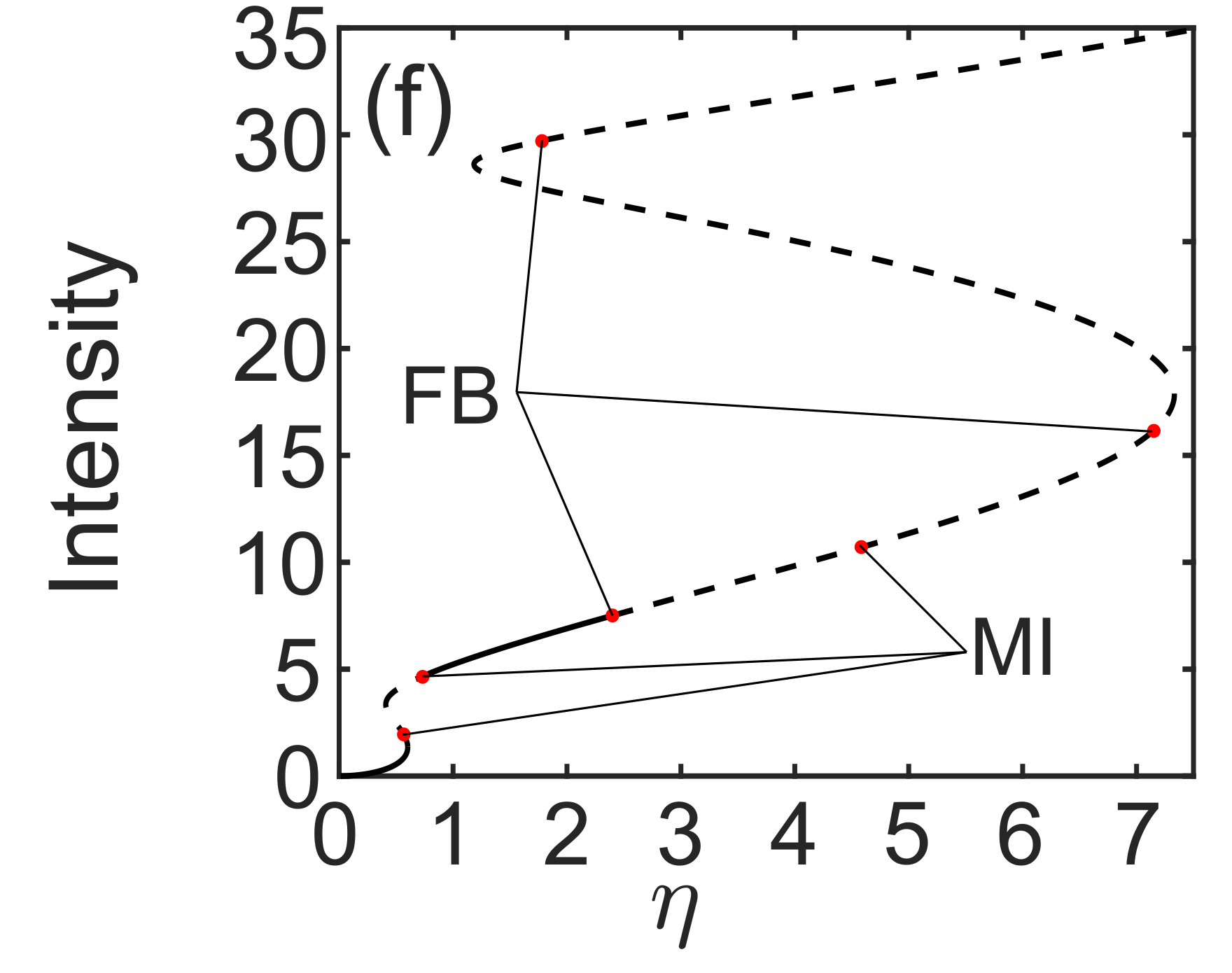}

\caption{S-shaped dependence of the intensity $I$ of CW solutions on the injection
rate $\eta$ (a). Curves of pseudo-continuous spectrum for different
values of $I$ (b) $I\approx1.7032$, (c) $I\approx2.4643$, (d) $I\approx4.0221$.
Saddle-node bifurcations (black curves) and modulational instability
(red curve) around a single cavity resonance on the ($\theta$, $\eta^{2}$)-plane
(e). CW solutions from (a) in the region of larger $\eta$ and $I$
(f). Parameters are: $\epsilon=0.5$, $\kappa=e^{-2\epsilon^{2}}$,
$\alpha=1\epsilon^{2}$, $\theta=-3.5\epsilon^{2}$, $a=1\epsilon$,
$b=1\epsilon^{2}$. In panels (a) and (f) unstable solutions are shown
by dashed lines. MI (FB) indicates the modulational (flip) bifurcation
points. \label{fig:cw} }
\end{figure}

S-shaped CW curve of Eq. (\ref{eq:Model}) is shown in Fig. \ref{fig:cw}(a).
It is seen that modulational instability takes place on the unstable
middle part of the CW curve and destabilizes its upper part. Figures
\ref{fig:cw}(c)-(d) present the real parts of the pseudo-continuous
eigenvalues $\gamma\left(\mu\right)$ calculated at the points indicated
in \ref{fig:cw}(a). Figure \ref{fig:cw}(e) illustrates the location
of the modulational instability curve between two saddle-node bifurcations
around a single cavity resonance in ($\eta^{2},\theta$)-plane. Finally,
Fig. \ref{fig:cw}(f) illustrates a growing multiplicity of the solutions
of Eq. (\ref{eq:cwcond}) due to the overlap of resonances with increasing
the injection rate.

\begin{figure}
\includegraphics[width=0.5\linewidth]{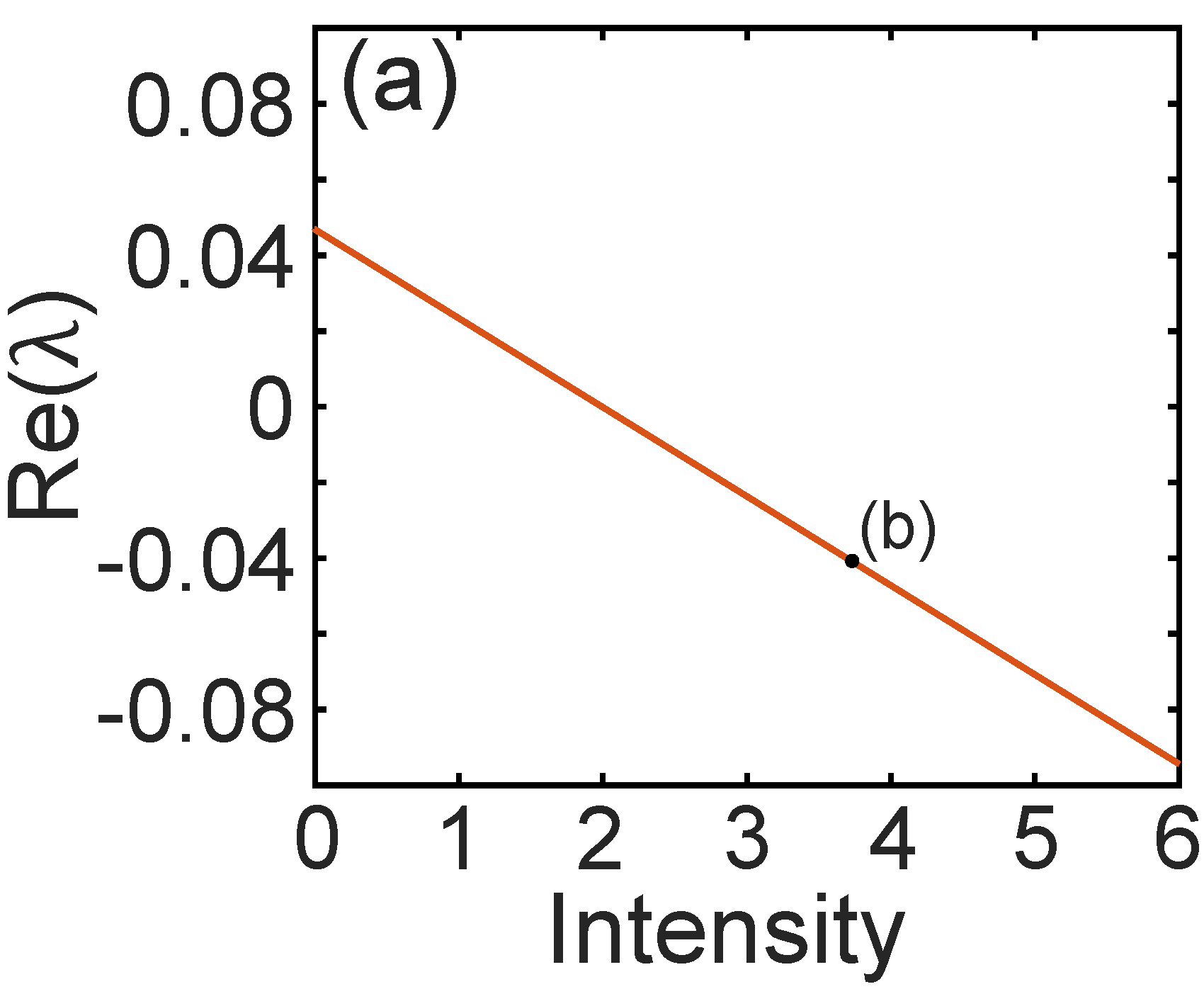}\includegraphics[width=0.5\linewidth]{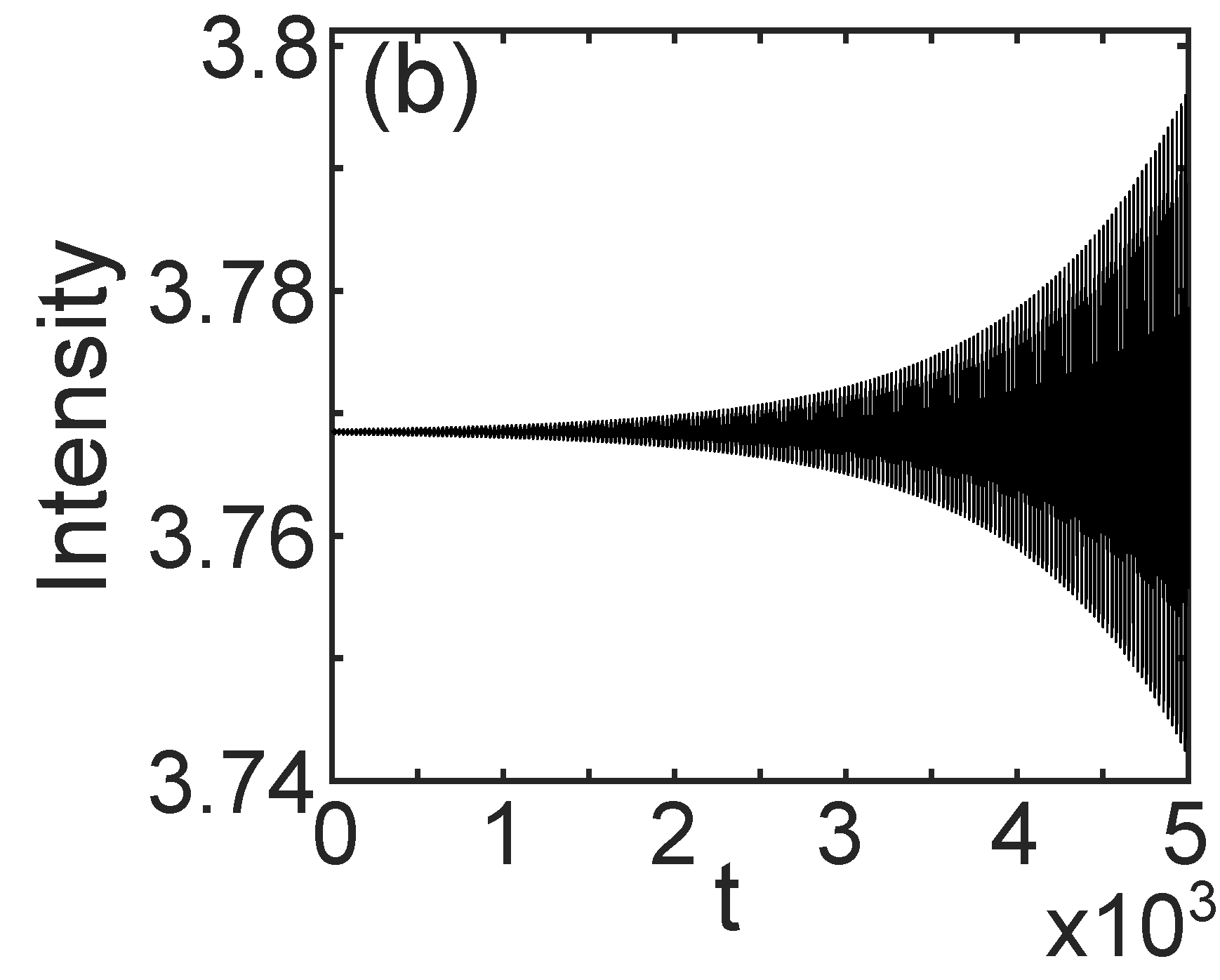}
\\

\caption{Largest real part of the discrete eigenvalues of the CW solution with
$b=(\sqrt{2}+0.01)a^{2}$ as a function of the CW intensity $I$ (a).
$a=0.1$. Evolution of the (modulationary unstable) state from the
upper bifurcation branch with $\eta=0.02$ (b). The solution diverges
with time. Other parameters are the same as in Fig. \ref{fig:cw}
except $\epsilon=0.1$. \label{fig:discrete2}}
\end{figure}

It is important also to check the discrete spectrum of nonlinear CW
solutions. The discrete spectrum is defined by the instantaneous part
of the model equation and can be obtained by solving the equation
$c_{0}=0$ with respect to $\lambda$, where $c_{0}$ is defined by
Eqs. (\ref{Y_eig}) and (\ref{eq:c2-0}). It is seen from Fig. \ref{fig:discrete2}
corresponding to $b=(\sqrt{2}+0.01)a^{2}$ that when $b$ exceeds
slightly the critical value defined by (\ref{eq:ineq}) the lower
part of the CW branch can become unstable with respect to the discrete
spectrum, see Fig. \ref{fig:discrete2}(a). The increase of the CW
intensity $I$ has a stabilizing effect on the discrete spectrum.
However, as it is seen from Fig. \ref{fig:discrete2}(b) the solution
of Eq. (\ref{eq:Model}) starting from the upper CW state becomes
unbounded in the limit $t\to\infty$. This behavior might be attributed
to a spurious instability.

\section{Cavity solitons}

\begin{figure}
\includegraphics[width=0.5\linewidth]{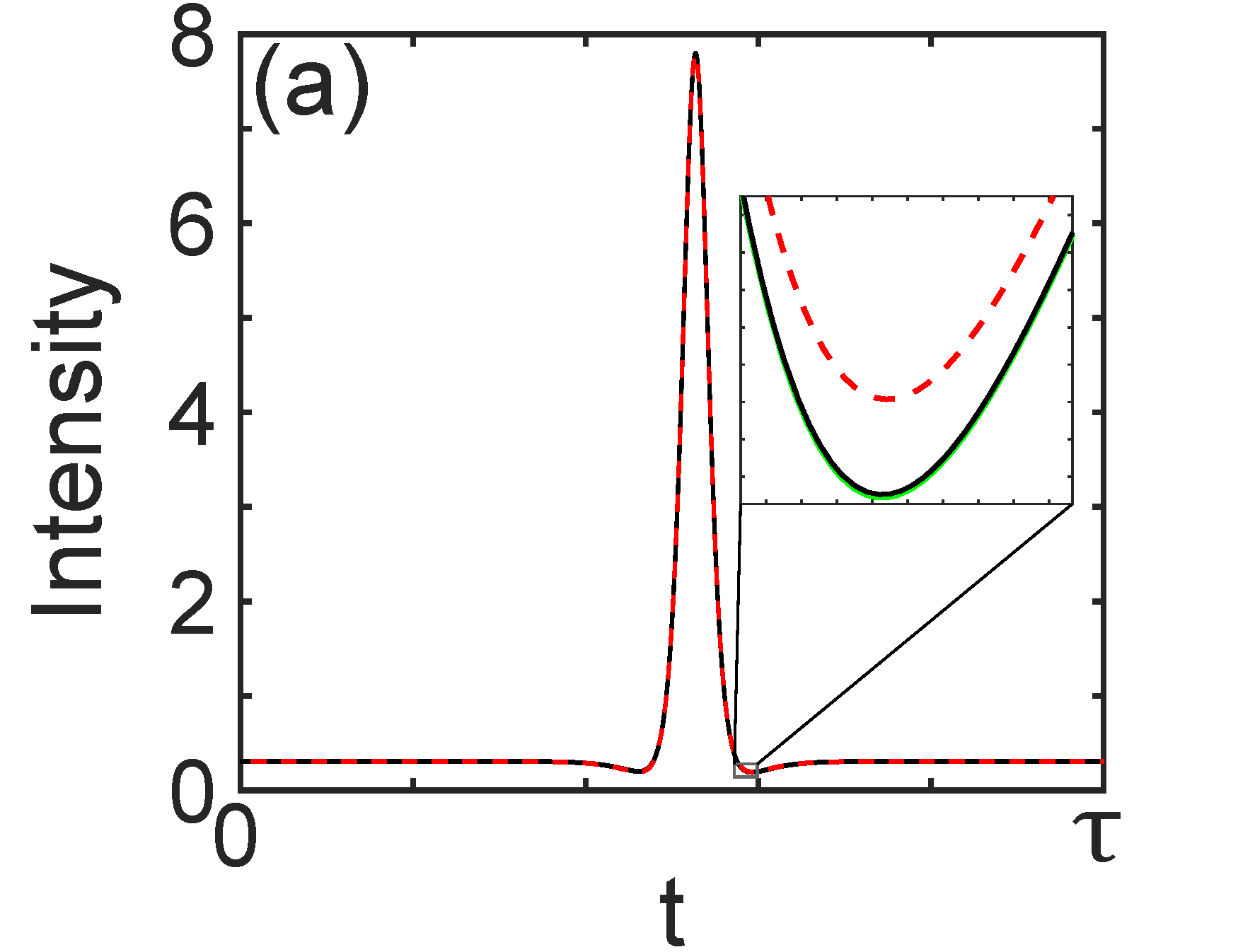}\includegraphics[width=0.5\linewidth]{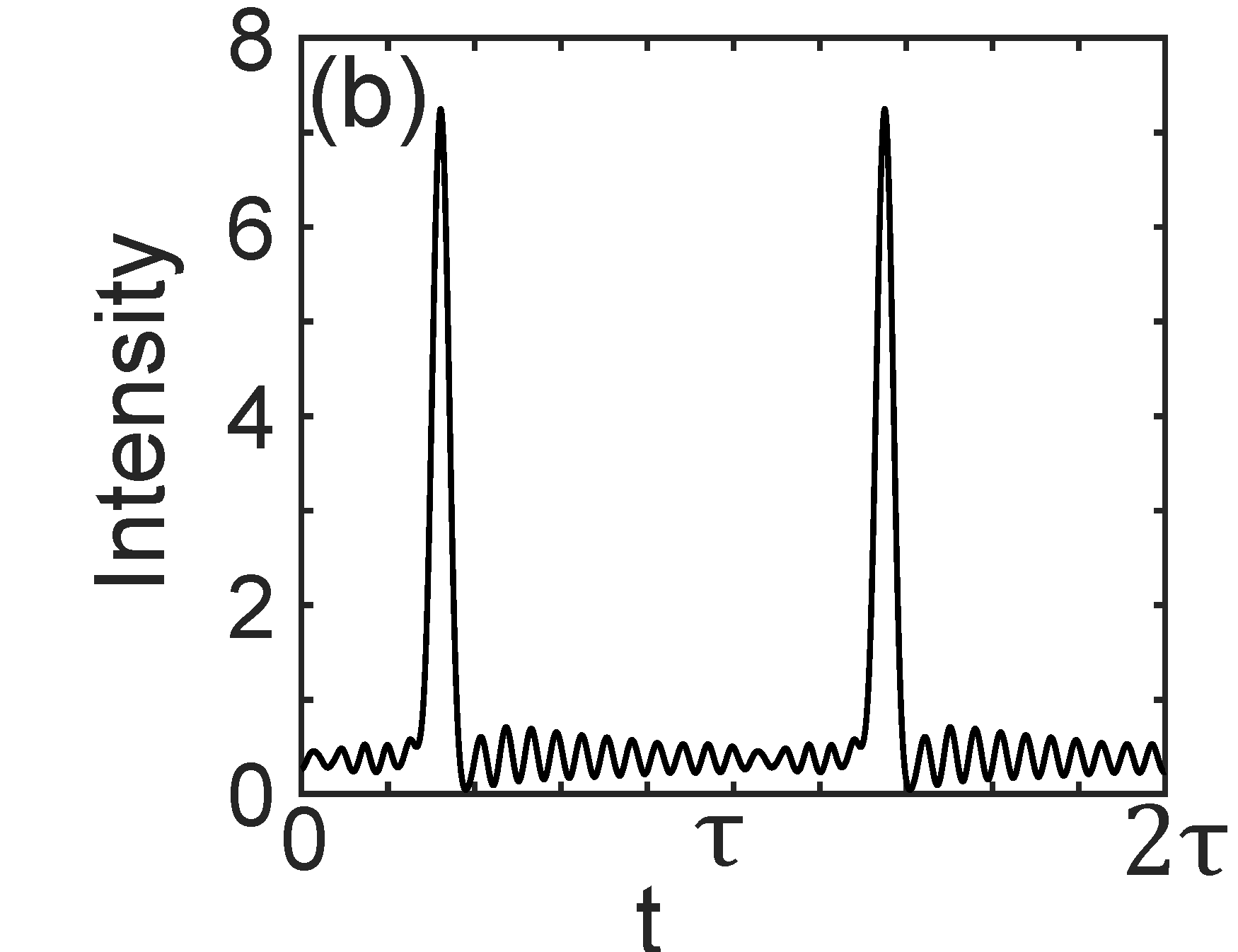}
\\
 \includegraphics[width=0.5\linewidth]{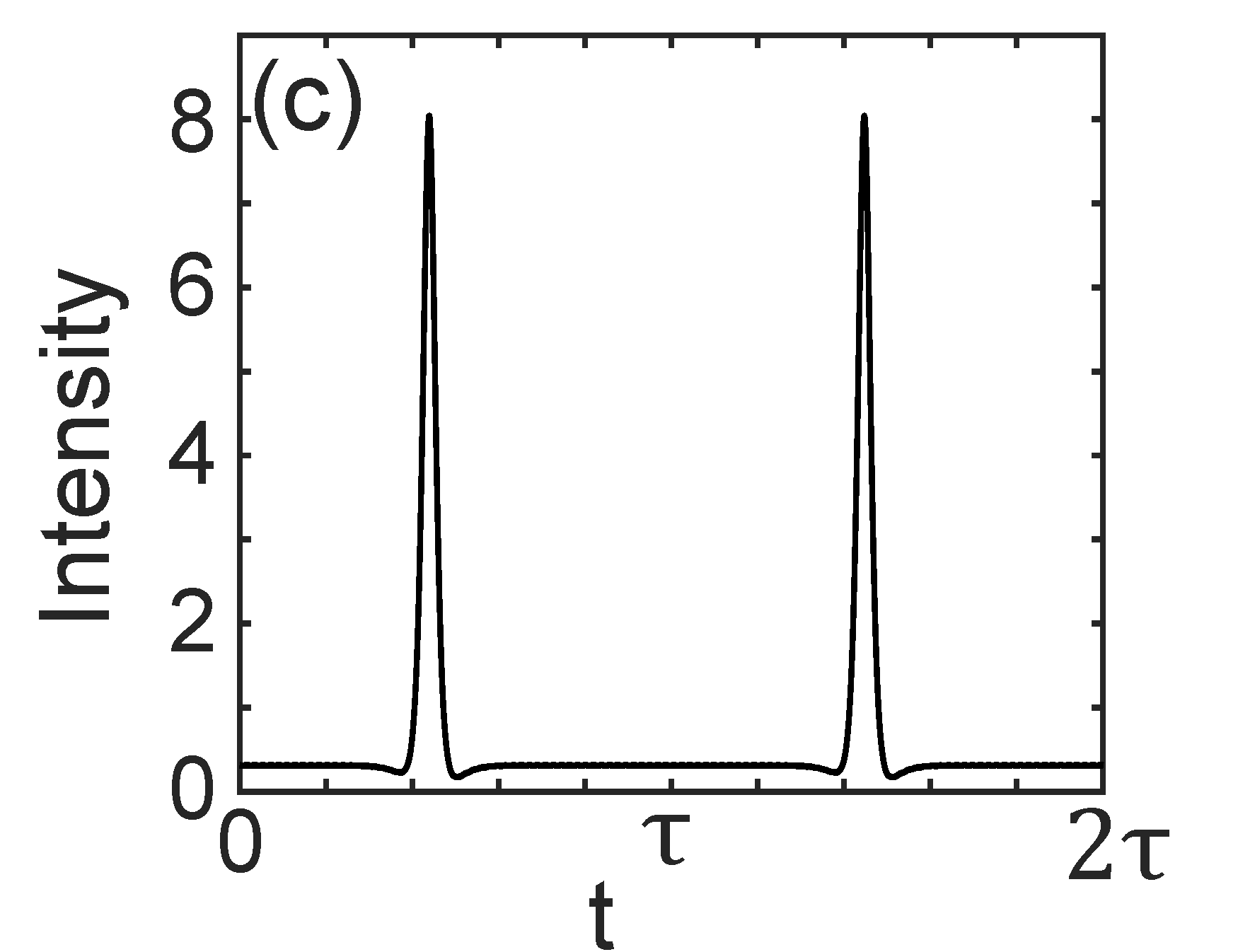}\includegraphics[width=0.5\linewidth]{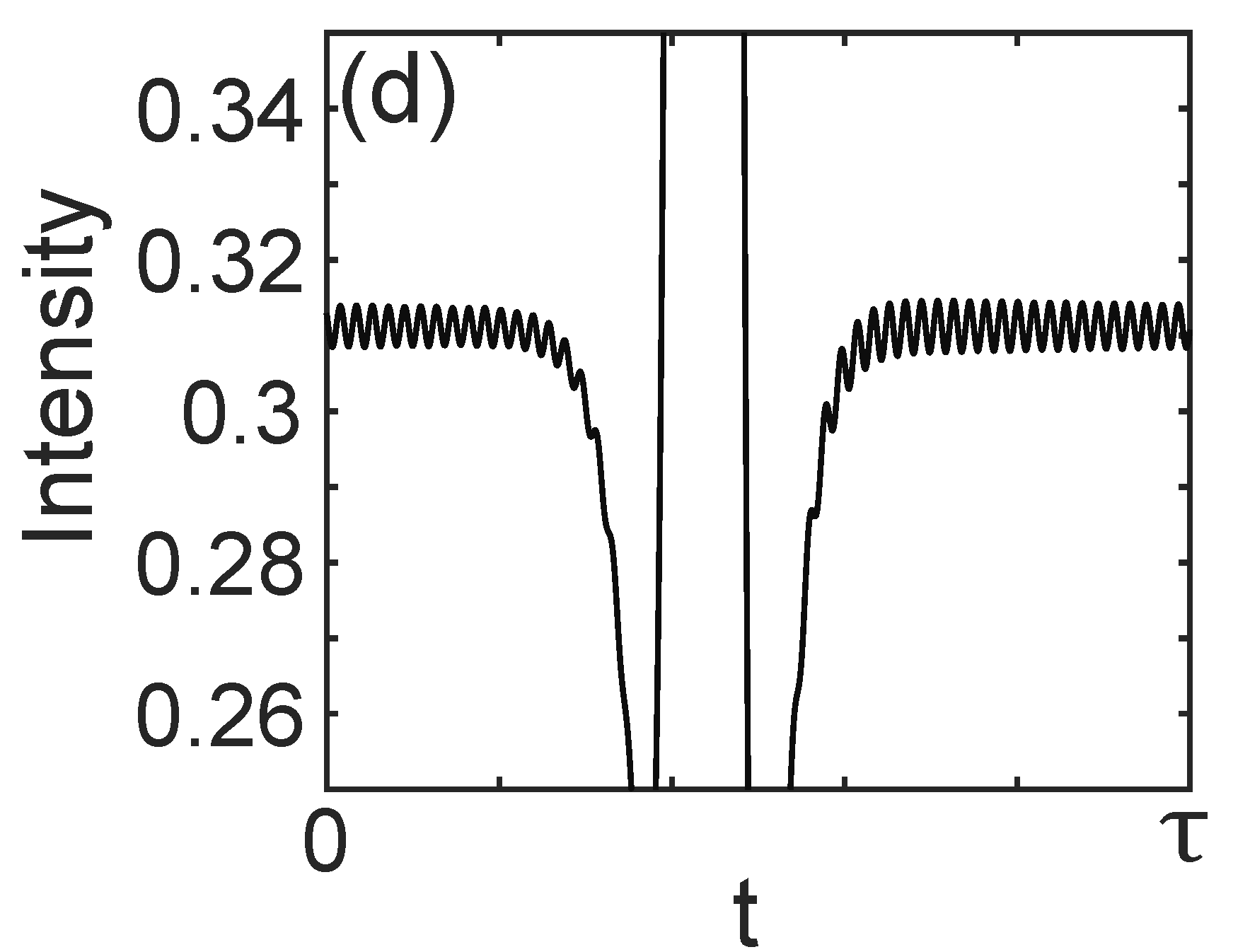}

\caption{TCS profiles calculated using the model equations (\ref{eq:Model})
(black line) with $\epsilon=0.02$ (a), $\epsilon=0.34$ (b), $\epsilon=0.1$
(c) and (d). Other parameters are given in the text. One can see from
panel (d) that already for $\epsilon=0.1$ the TCS tails exhibit a
Cherenkov radiation, which is more pronounced at the trailing tail.
In panel (a) the TCS of Eq. (\ref{eq:Model-2}) is shown by green
line and the soliton of the LLE (\ref{eq:LLE-standard}) - by red
dashed line. Parameters are: $\kappa=e^{-2\epsilon^{2}}$, $\alpha=\epsilon^{2}$,
$a=\epsilon$, $b=\epsilon^{2}$, $\theta=-3.5\epsilon^{2}$, $\eta=1.855\epsilon^{2}$.\label{fig:sols}}
\end{figure}

In this section we investigate numerically TCS solutions of the NDDE
models using the RADAR5 code written in FORTRAN \cite{guglielmi2001implementing}.
We start with the delay time $\tau=25$ and the parameter values close
to the LLE limit defined by Eqs. (\ref{eq:LLE_limit1}) and (\ref{eq:LLE_limit2})
, $\kappa=e^{-2\epsilon^{2}}$, $\alpha=\epsilon^{2}$, $a=\epsilon$,
$b=\epsilon^{2}$, $\theta=-3.5\epsilon^{2}$, $\eta=1.855\epsilon^{2}$
with $\epsilon=0.02$. The calculated TCS solution of Eq. (\ref{eq:Model})
is shown in Fig. \ref{fig:sols}(a) by black line. This soliton is
very close to the TCS of the model (\ref{eq:Model-2}) shown by green
line. Dissipative soliton of the LLE (\ref{eq:LLE-standard}) is indicated
by red dashed line in the same figure for the LLE parameter values
obtained using the relations (\ref{eq:LLE_limit1}) and (\ref{eq:LLE_limit2}),
$k=1$, $\Theta=-3.5$, $\chi=1$, $b=1$, $r=1.855$ in Eq. (\ref{eq:LLE-standard}).
With increasing $\epsilon$ the soliton profile gets more asymmetric,
see Figs.\ref{fig:sols}(c) and (d) obtained with $\epsilon=0.1$.
As it is seen in Fig. (d), the soliton tails exhibit slowly decaying
oscillations, which can be attributed to the so-called Cherenkov radiation
induced by high order dispersion \cite{karpman1993radiation,Akhmediev95}.
The Cherenkov radiation amplitude increases with $\epsilon$, see
Fig. \ref{fig:sols}(b) obtained with $\epsilon=0.34$ and eventually
destroys the TCS.

\begin{figure}
\centering \includegraphics[width=0.7\linewidth]{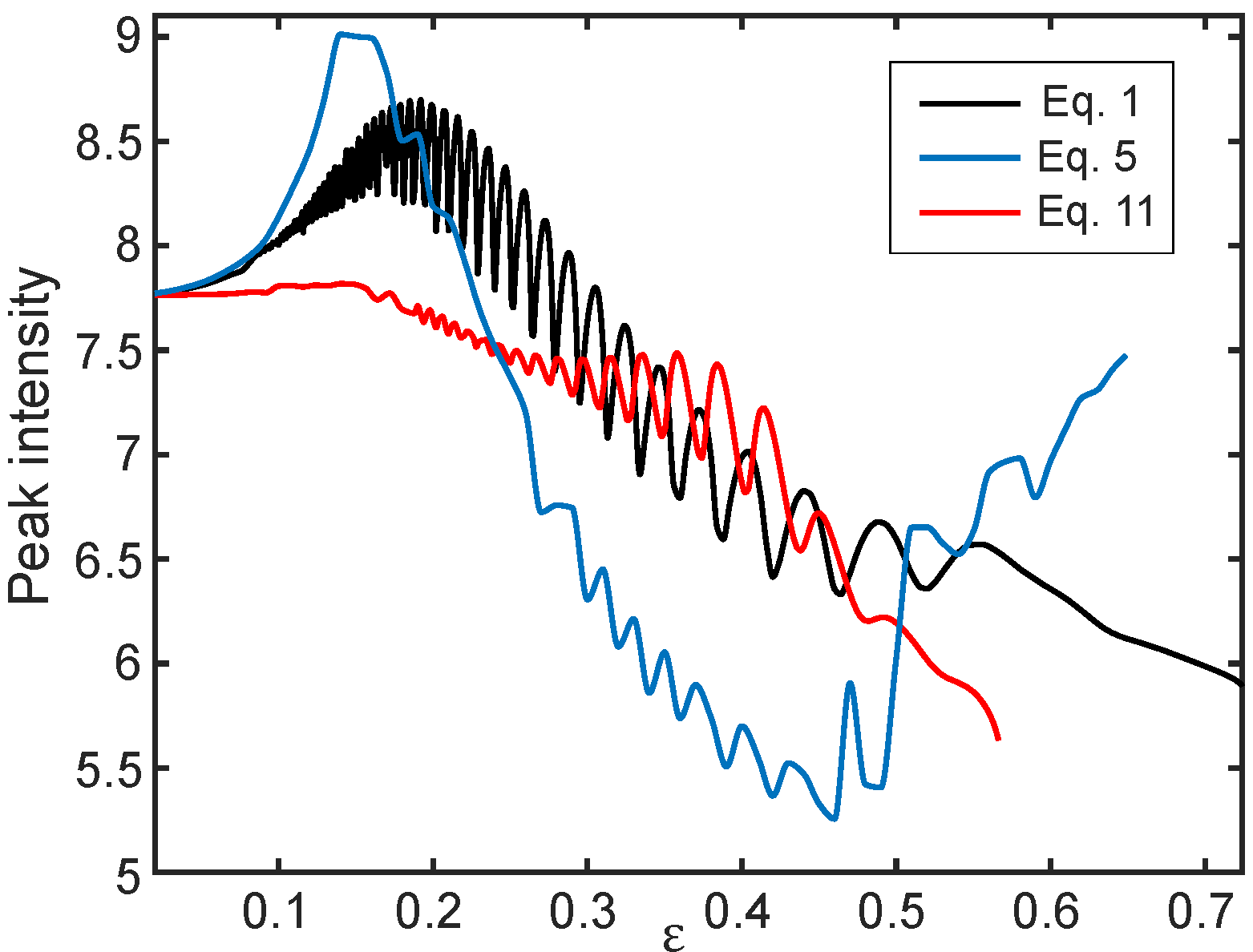}

\caption{TCS peak power as a function of the parameter $\epsilon$ calculated
using Eq. (\ref{eq:Model}) (black), Eq. (\ref{eq:Model-2}) (blue),
and Eq. (\ref{eq:Model-mean_field}) (red) with $\eta=1.855\epsilon^{2}$
and $\tau=25$. Other parameters are the same as in Fig.\ref{fig:sols}.\label{fig:I_eps}}
\end{figure}

The dependence of the TCS peak power on the parameter $\epsilon$
calculated using the models (\ref{eq:Model}) (black line) and (\ref{eq:Model-2})
(blue line) with exponential nonlinearity, and the model (\ref{eq:Model-mean_field})
(red line) with cubic nonlinearity is shown in Fig. \ref{fig:I_eps}.
It is seen that while sufficiently close to the LLE limit, $\epsilon\to0$,
this dependence looks rather smooth, it becomes fast oscillating with
the increase of $\epsilon$. Such oscillatory behavior can be explained
by the interaction of a TCS with its own tails leading either to constructive
or destructive interference due to the presence of Cherenkov radiation.
Furthermore, it is seen that the oscillation frequency increases with
the decrease of $\epsilon$, which is in agreement with the fact that
the frequency of the Cherenkov radiation tends to infinity when the
third and/or fourth order dispersion coefficients tend to zero, see
e. g. \cite{vladimirov2018effect,vladimirov2021dissipative}. Note
that the more exact model (\ref{eq:Model-2}) demonstrates weaker
dependence of the TCS peak power on the Cherenkov radiation than Eq.
(\ref{eq:Model}). Therefore, Eq. (\ref{eq:Model-2}) may be more
suitable for the description of the system dynamics far away from
the LLE limit.

\begin{figure}
\centering \includegraphics[width=1\linewidth]{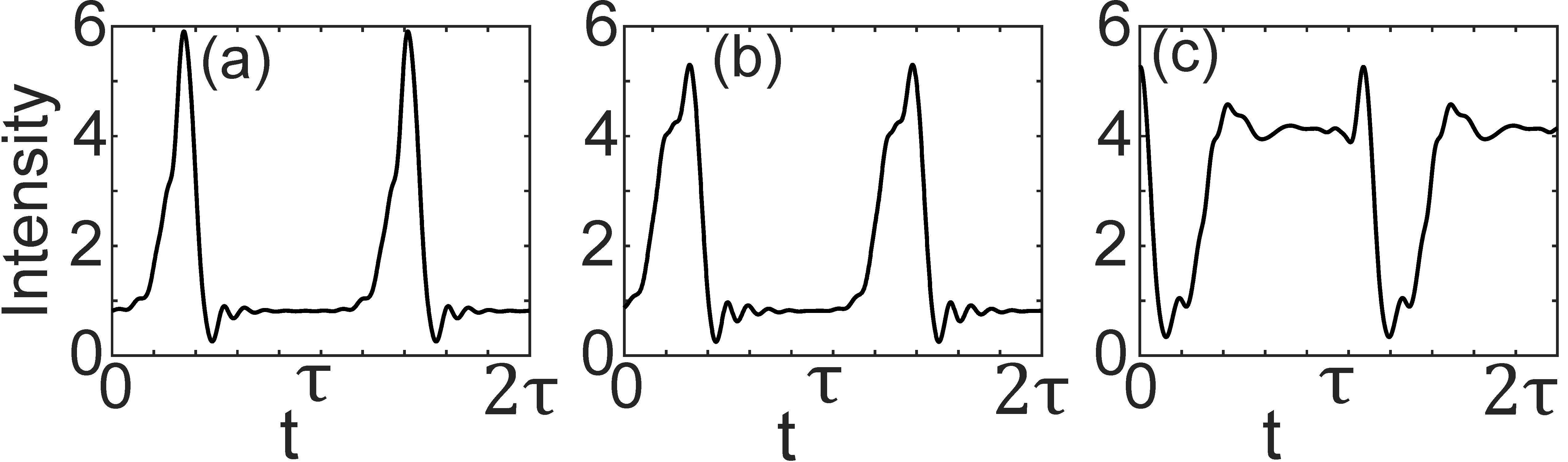}

\caption{Different types of localizes solutions of the NDDE model (\ref{eq:Model})
calculated for $\epsilon=0.722$. The TCS shown in panel (a) corresponds
to the black curve shown in Fig.\ref{fig:I_eps}. Other parameters
are the same as in Fig.\ref{fig:I_eps}. \label{fig:dif_sols}}
\end{figure}

At relatively high values of $\epsilon$ different types of TCS can
appear, as it is shown in Fig. \ref{fig:dif_sols}. Their stability
ranges in the parameter space are relatively small, but with $\epsilon=0.722$
can coexist for the same set of parameters.

\begin{figure}
\includegraphics[width=0.5\linewidth]{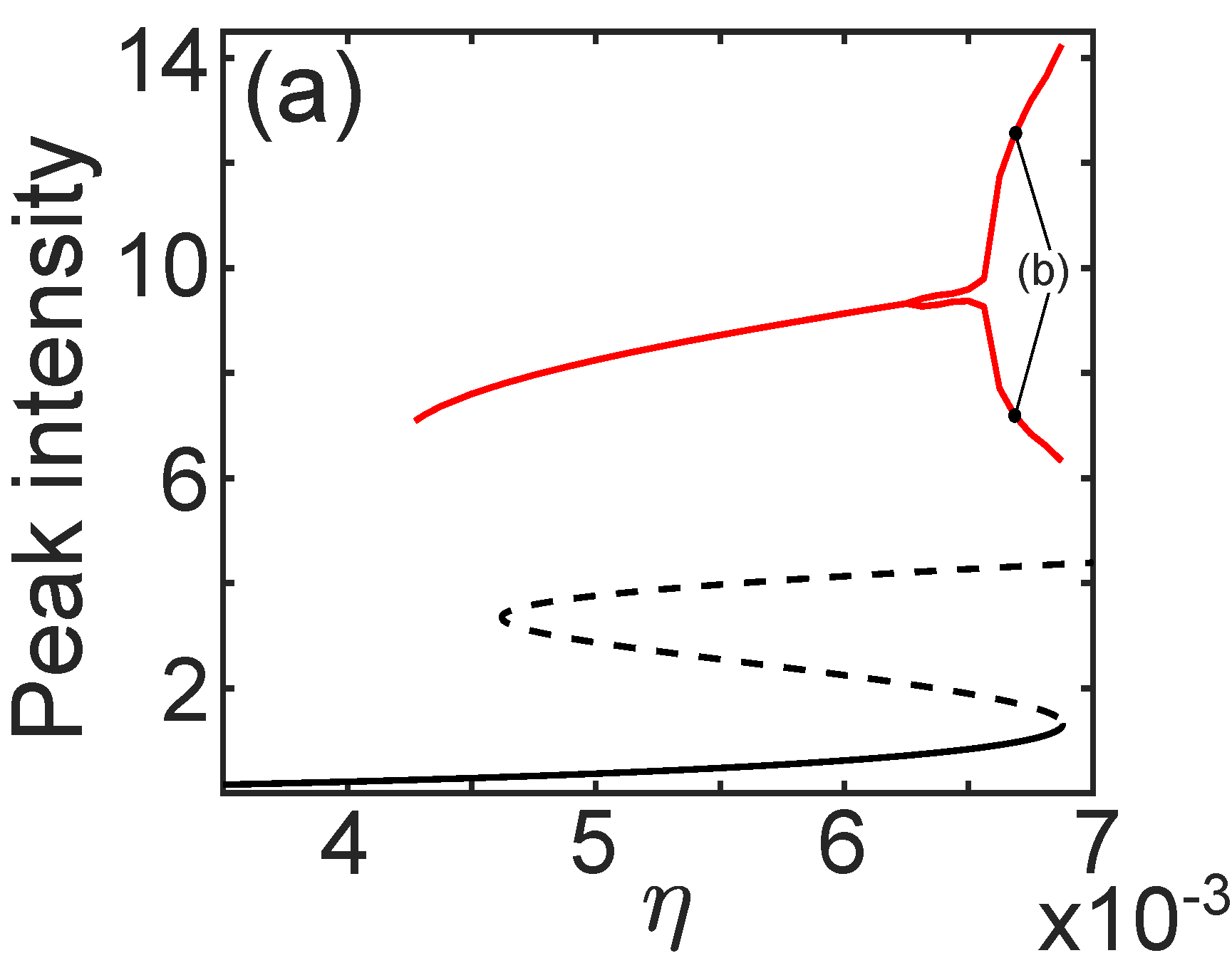}\includegraphics[width=0.5\linewidth]{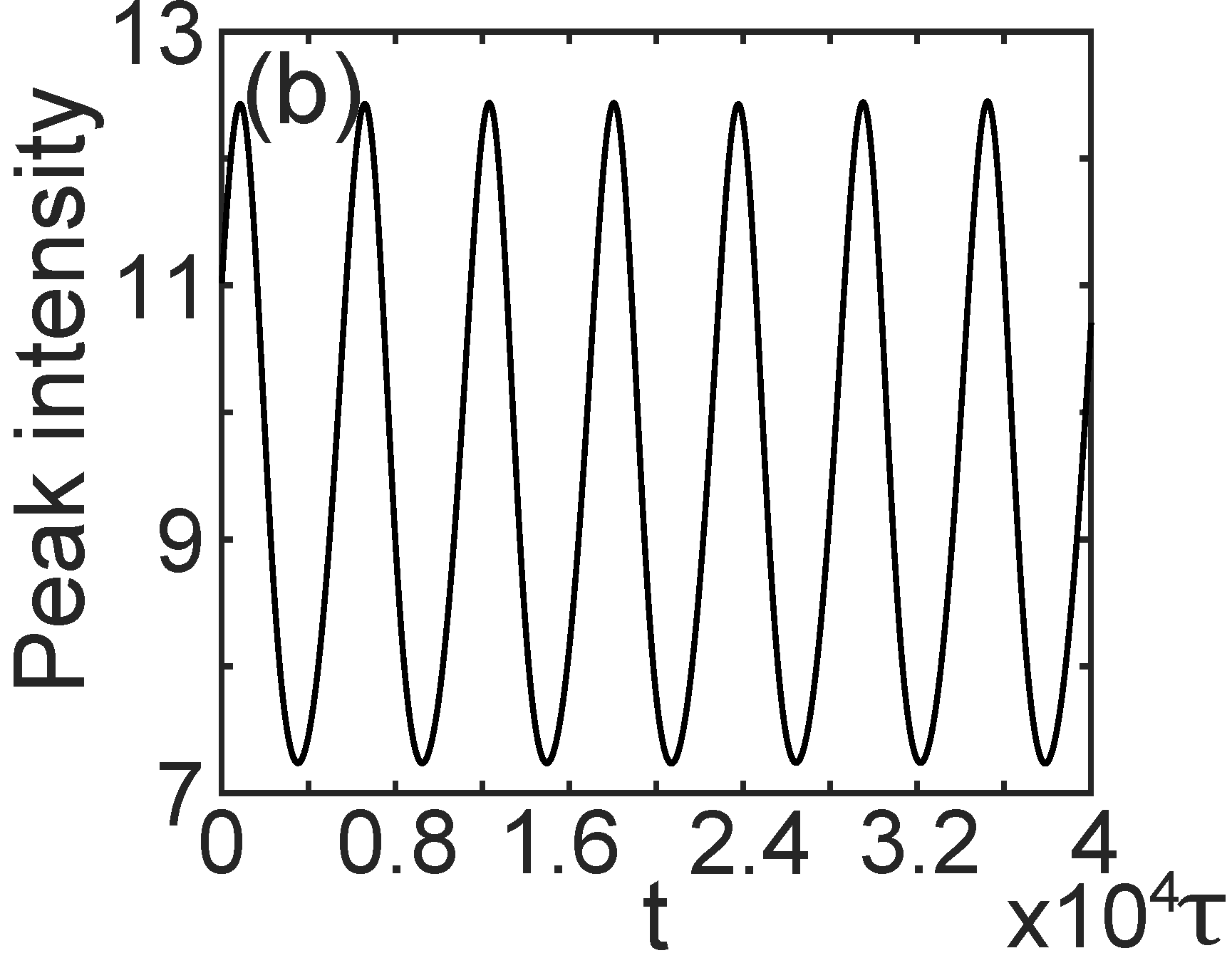}

\caption{TCS peak power (red curve) and intensity of CW solution (black curve)
as functions of the injection rate $\eta$ (a). The TCS starts to
oscillate above the Andronov-Hopf bifurcation threshold at $\eta\approx6.25\times10^{-3}$.
Red lines show maximal and minimal peak power within the oscillation
period. Time-trace of the TCS peak power calculated for $\eta=6.7\times10^{-3}$
(b). $\epsilon=0.05$. Other parameters are the same as in Fig.\ref{fig:I_eps}
\label{fig:oscill}}
\end{figure}

It is known that the soliton of the LLE can undergo an oscillatory
instability with the increase of the injection rate. The NDDE model
(\ref{eq:Model}) also shows a similar behavior. Figure \ref{fig:oscill}
illustrates the appearance of undamped oscillations of the soliton
peak power for fixed relatively small $\epsilon=0.05$. At higher
values of $\epsilon$ there are different types of oscillating TCSs.

\section{Conclusion}

To conclude, we have developed a second order NDDE model of a ring
dispersive Kerr cavity with a coherent optical injection. Similarly
to the first order NDDE model discussed in \cite{seidel2022conservative}
in the non-dissipative limit this model is reversible and has a conserved
quantity. In a certain parameter range and under the mean field and
large delay approximations the NDDE model can be reduced to the famous
LLE model. However, unlike the LLE and similarly to the Ikeda map
\cite{haelterman1992dissipative} and generalized LLE \cite{kartashov2017multistability}
Kerr cavity models, the NDDE model is able to describe the overlap
of the resonances associated with different cavity modes. We have
shown that TCSs can exist in the NDDE model not only close to the
LLE limit, but also beyond this limit. In the latter case they are
strongly affected by the Cherenkov radiation, which is induced by
high order dispersion and eventually destroys the TCS. An important
advantage of the NDDE model is that it can be analyzed numerically
using standard codes, such as RADAR5 \cite{guglielmi2001implementing}
and DDE-biftool \cite{DDEbiftool}. Furthermore, after appropriate
modifications this model can be applied to study the effect of chromatic
dispersion on the dynamics of mode-locked lasers and other laser systems.
The NDDE model might be also useful for the consideration of the coupled-cavity
systems, such as an optical microcavity pumped by a semiconductor
mode-locked laser modeled by the DDE mode-locking model \cite{VT05,vladimirov,VT04}.
This model can be easily extended by including higher order dispersion
terms into it. 
\begin{acknowledgments}
Stimulating discussions with Guillaume Huyet, Dmitry Turaev, Sergei
Turitsyn, and Matthias Wolfrum, as well as the support by the Deutsche
Forschungsgemeinschaft (DFG projects No. 445430311 and No. 491234846)
are gratefully acknowledged. 
\end{acknowledgments}

\bibliographystyle{apsrev4-2}

\end{document}